\documentclass[twocolumn,prb,aps,showpacs]{revtex4}
\usepackage{graphicx}
\usepackage{amsmath}
\usepackage{bm}

\begin{document}

\title{Design of electron wave filters in monolayer graphene by tunable transmission gap}

\author{Xi Chen$^{1,2}$\footnote{Email address: xchen@shu.edu.cn}}

\author{Jia-Wei Tao$^{1}$}

\affiliation{$^{1}$ Department of Physics, Shanghai University,
200444 Shanghai, People's Republic of China}

\affiliation{$^{2}$ Departamento de Qu\'{\i}mica-F\'{\i}sica,
UPV-EHU, Apdo 644, 48080 Bilbao, Spain}

\begin{abstract}
We have investigated the transmission in monolayer graphene barrier
at non-zero angle of incidence. Taking the influence of parallel
wave vector into account, the transmission as the function of
incidence energy has a gap due to the evanescent waves in two cases
of Klein tunneling and classical motion. The modulation of the
transmission gap by the incidence angle, the height and width of
potential barrier may lead to potential applications in
graphene-based electronic devices.



\end{abstract}

\maketitle

Graphene has become a subject of intense interest
\cite{Neto-GPN,Beenakker} since the graphitic sheet of one-atom
thickness has been experimentally realized by A. K. Geim \textit{et
al.} in 2004 \cite{Novoselov-GMJ}. The valence electron dynamics in
such a truly two-dimensional (2D) material is governed by a massless
Dirac equation. As a result, graphene exhibits many unique
electronic and transport properties \cite{Novoselov,Katsnelsona-N},
such as the half-integer quantum Hall effect
\cite{Novoselov-GMJK,Zhang-TS,Gusynin-S} and the minimum
conductivity \cite{Novoselov-GMJK}. Furthermore, another one is the
perfect transmission, in particular, for normal incidence, through
arbitrarily high and wide graphene barriers, which is referred to as
Klein tunneling \cite{Katsnelson-NG}. All these properties are
significant in the design of various graphene-based devices, and
graphene is thus regarded as a perspective base for the post-silicon
electronics.

Until recently, the transport properties of massless Dirac fermions,
including Klein tunneling and resonance transmission, have been
extensively studied in the single or double graphene barriers and
graphene superlattices \cite{Katsnelson-NG, Pereira,Pereira-VP,Bai}.
However, inhomogeneous magnetic fields on the nanometer scale has
been lately suggested to circumvent Klein tunneling and produce
confined graphene-based structures \cite{Martino,Masir-VMP,Zhai}. It
was found that the angular range of the transmission through
monolayer or bilayer graphene with magnetic barrier structures could
be efficiently controlled and resulted in the direction-dependent
wave vector filter \cite{Masir-APL,Masir-PRB}.

In this Letter, we will investigate that the transmission of
Dirac-like electrons in 2D monolayer graphene barrier at non-zero
incidence angle. It is shown that when the electrons are obliquely
incident on the potential barrier, the transmission has a gap, which
depends strongly on the incidence angle, the width and height of
barrier. This tunable transmission gap is quite different from the
perfect transparency for the normal incidence \cite{Katsnelson-NG}
and does result from evanescent waves in two cases of Klein
tunneling and classical motion due to the influence of parallel wave
vector. In fact, based on the mechanism of Klein tunneling, a single
graphene barrier is equivalent to a more complicated resonant
tunneling device in common semiconductor heterostructures, at least
from the point of view of the transmission \cite{Dragoman-APL}.
Thus, these phenomena will provide a completely different mechanism
of electron wave filters at the nanoscale level with more
flexibility and simplicity in design than those in multiple
semiconductor quantum wells \cite{Garg}.

\begin{figure}[]
\scalebox{0.80}[0.80]{\includegraphics{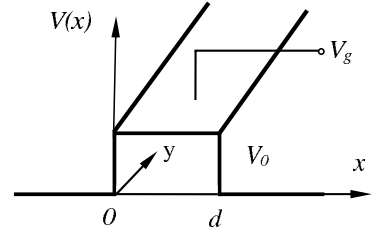}}\caption{Schematic
diagram for 2D graphene barrier.\label{fig.1}}
\end{figure}

Consider the ballistic electrons with Fermi energy $E$ at angle
$\phi$ with respective to the $x$ axis incident upon a 2D potential
barrier, as shown in Fig. \ref{fig.1}, where the tunable potential
barrier is formed by a bipolar junction ($p$-$n$-$p$) within a
single-layer graphene sheet with top gate voltage $V_g$
\cite{Huard}, $V_0$ and $d$ are the height and width of potential
barrier, respectively. Since graphene is a 2D zero-gap semiconductor
with the linear dispersion relation, $E= \hbar k_{F}\upsilon_{F}$,
the electrons are formally described by the Dirac-like hamiltonian
\cite{Katsnelson-NG}, $\hat{H_{0}}=-i\hbar v_{F}\sigma\nabla $,
where $v_F \approx 10^{6} m \cdot s^{-1}$ is the Fermi velocity,
$k_F$ is the Fermi wave vector, and $\sigma=(\sigma_{x}, \sigma_{y})
$ are the Pauli matrices. The wave function of the incident
electrons is assumed to be
\begin{equation}
\Psi_{in}(x,y)=
   \left(\begin{array}{c}
      1 \\
      se^{i\phi} \\
   \end{array}\right)
e^{i(k_{x}x+k_{y}y)},
\end{equation}
the wave function of the transmitted one can be thus expressed by
\begin{equation}
\Psi_{t}(x,y)=t
 \left(\begin{array}{c}
      1 \\
      se^{i\phi} \\
   \end{array}\right)
e^{i(k_{x}x+k_{y}y)},
\end{equation}
where $k_{x}=k_{F}\cos\phi$ and $k_{y}=k_{F}\sin\phi$ are the
perpendicular and parallel wave vector components outside the
barrier, $k'_F=|E-V_{0}|/\hbar v_{F}$,
$q_{x}=(k^{'2}_{F}-k_{y}^{2})^{1/2}$, $\theta=\arctan(k_{y}/q_{x})$,
$s=sgn(E)$ and $s'=sgn(E-V_{0})$. According to the boundary
conditions, the transmission coefficient is determined by
\begin{equation}
\label{transmission} t=\frac{2ss'\cos\phi\cos\theta}
{ss'[e^{-iq_{x}d}\cos(\phi+\theta)+e^{iq_{x}d}\cos(\phi-\theta)]-2i\sin(q_{x}d)}.
\end{equation}
In what follows we will discuss the transmission and reflection in
two different cases of Klein tunneling ($E<V_0$) and classical
motion ($E>V_0$).

\textit{Case 1}: Klein tunneling ($ss'=-1$). The transmission
probability $T$ can be given by Eq. (\ref{transmission}),
\begin{equation}
T\equiv|t|^2=\left[\cos^{2}(q_{x}d)+\frac{(k^{2}_{y}+k_{F}k^{'}_{F})^{2}}{k^{2}_{x}q^{2}_{x}}\sin^{2}(q_{x}d)\right]^{-1}.
\end{equation}
Under the resonance conditions, $q_x d = N \pi$, ($N=0, 1,...$), the
transmission probability $T$ is equal to $1$. In addition, the
barrier always remains perfectly transparent at the normal incidence
$\phi=0$ \cite{Katsnelson-NG}, which is so-called Klein paradox in
QED \cite{Dombey}. However, the transmission can be divided into
evanescent and propagating modes, taking the influence of the
parallel wave vector $k_y$ into account. The electrons can tunnel
through the potential barrier when $\phi > \phi_c$, where the
critical angle for total reflection can be defined as
\begin{equation}
\label{critical angle} \phi_c = \sin^{-1}
\left(\frac{V_0}{E}-1\right),
\end{equation}
with the necessary condition $E<V_0<2E$. In this case, the
transmission probability damped exponentially in the following form:
\begin{equation}
\label{decay-1}
T \approx  \frac{4
k^{2}_{x}q^{2}_{x}}{k^{2}_{x}q^{2}_{x} +
(k^{2}_{y}+k_{F}k^{'}_{F})^{2}} e^{- 2 \kappa d},
\end{equation}
where $\kappa = [k^2_y -(E-V_0)^2/\hbar^2 v^2_F]^{1/2}$ is the decay
constant. As a matter of fact, the electrons can traverse through
the potential barrier in propagating mode at any incidence angles,
when the critical angle $\phi_c$ is no longer valid for $V_0>2E$
\cite{Katsnelson-NG}. These results presented here can offer the
complementary understanding of the evidence against Klein paradox in
graphene \cite{Bowen,Dragoman}.

\begin{center}
\begin{figure}[ht]
\scalebox{0.36}[0.36]{\includegraphics{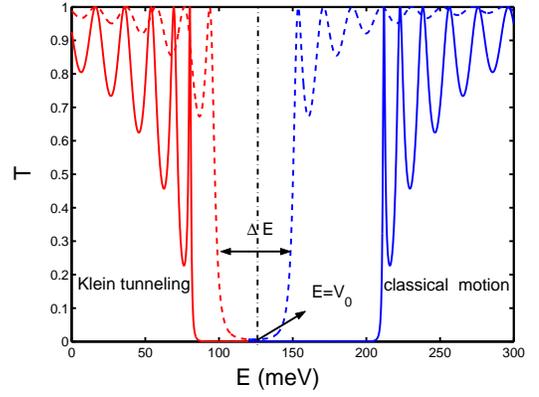}} \caption{(Color
online) Transmission gap as the function of the incident energy $E$,
where $d=100nm$, $V_0=120 meV$, solid and dashed curves correspond
to $\phi=25^\circ$ and $10^\circ$, respectively. \label{fig.2}}
\end{figure}
\end{center}

\begin{figure}[]
\scalebox{0.35}[0.35]{\includegraphics{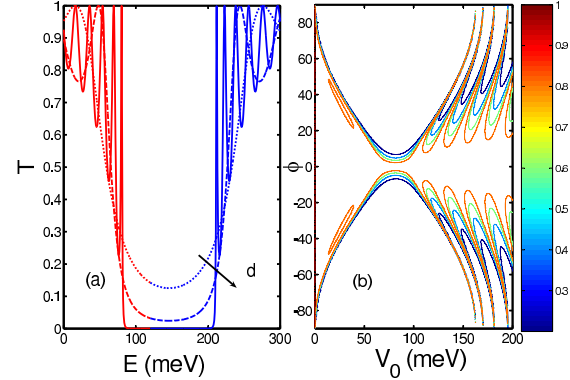}} \caption{(Color
online) Dependence of transmission gap on the width and the height
of potential barrier, where (a) $V_0=120 meV$, $\phi=25^\circ$,
solid, dashed and dotted curves correspond to $d=100 nm$, $30 nm$,
and $20 nm$, (b) $E=80 meV$ and $d=100 nm$. \label{fig.3}}
\end{figure}
\textit{Case 2}: Classical motion ($ss'=1$). The transmission
probability can be rewritten by
\begin{equation}
T=\left[\cos^{2}(q_{x}d)+\frac{(k^{2}_{y}-k_{F}k^{'}_{F})^{2}}{k^{2}_{x}q^{2}_{x}}\sin^{2}(q_{x}d)\right]^{-1}.
\end{equation}
Similarly, when the incidence angle $\phi$ is less than the critical
angle,
\begin{equation}
\label{critical angle-2} \phi'_c = \sin^{-1}
\left(1-\frac{V_0}{E}\right),
\end{equation}
the transmission probability $T$ depends periodically on the width
$d$ of barrier. On the contrary, the electrons with the incidence
angle of $\phi
> \phi'_c$ tunnel through the potential barrier with
the transmission probability,
\begin{equation} \label{decay-2} T
\approx  \frac{4 k^{2}_{x}q^{2}_{x}}{k^{2}_{x}q^{2}_{x} +
(k^{2}_{y}-k_{F}k^{'}_{F})^{2}} e^{- 2 \kappa d}.
\end{equation}
Based on the properties in two cases of Klein tunneling and
classical motion, the transmission as the function of incidence
energy $E$ has a gap, as shown in Fig. \ref{fig.2}. Since $q^2_x =
(E-V_0)^2/\hbar^2 v^2_F-k^2_y <0$, the energy region of the
transmission gap is given by $ V_0-\hbar k_y v_F  < E < V_0 + \hbar
k_y v_F, $ which leads to the width of transmission gap as follows,
\begin{equation}
\Delta E = 2\hbar k_y v_F.
\end{equation}
Obviously, the transmission gap with the center $E=V_0$ becomes
narrower with the decrease of the incidence angle, and even vanishes
at normal incidence. The transmission gap is due to the evanescent
waves in two cases of Klein tunneling and classical motion, so it
has nothing to do with magnetic barriers in graphene
\cite{Masir-APL,Masir-PRB}, and is also quite different from that in
graphene double barriers \cite{Pereira-VP} or superlattices
\cite{Bai}, where the interference plays an important role in the
transmission resonances and related transmission gap.

\begin{figure}[]
\scalebox{0.35}[0.35]{\includegraphics{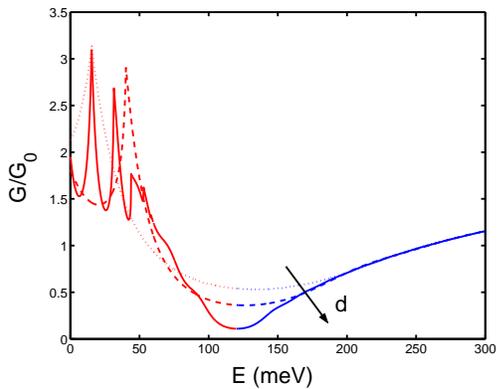}} \caption{(Color
online) Conductance $G$ as the function of incident energy, where
the parameters are the same as in Fig. \ref{fig.3} (a).}
\label{fig.4}
\end{figure}
Fig. \ref{fig.3} (a) further indicates the dependence of
transmission gap on the width of potential barrier. It is indicated
that the transmission gap will become deeper with the increase of
the barrier width, due to the decrease of the decay factor $\exp{(-
2 \kappa d)}$ in Eqs. (\ref{decay-1}) and (\ref{decay-2}). The
dependence of the transmission gap on the height of potential
barrier and the incidence angle is also shown in Fig. \ref{fig.3}
(b). Interestingly, the center of the transmission gap can be
controlled by changing the barrier height or strength (e.g. via
adjusting a gate-voltage $V_g$ in tunable graphene potential
barrier) \cite{Huard}. That is to say, the incident energy can be
selected by the tunable transmission gap, which results in an
alternative way to realize an electron wave energy filter.
\begin{figure}[] \scalebox{0.35}[0.35]{\includegraphics{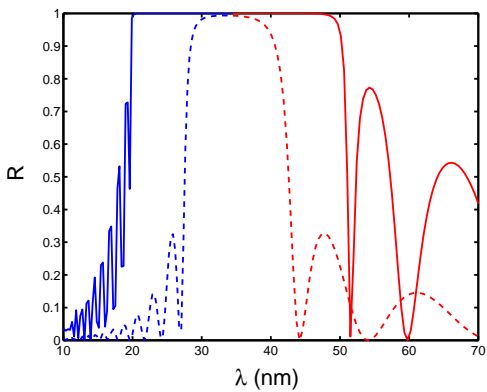}}
\caption{(Color online) Reflection probability $R$ as the function
of $\lambda$, where the parameters are the same as in Fig.
\ref{fig.2}. \label{fig.5}}
\end{figure}Moreover,
the transmission gap discussed here is also related to the negative
differential resistance \cite{Dragoman-APL}. In a word, the
transmission gap in actual device structure can result in various
graphene-based electronic devices.

In addition, the ballistic conductance under zero temperature is
calculated by electron flow averaged over the half of the Fermi
surface \cite{Matulis,Masir-PRB},
\begin{equation}
G=G_0 \int^{\pi/2}_{-\pi/2} T(E_F, E_F \sin\phi) \cos \phi d \phi,
\end{equation}
with Fermi energy $E_F$ and the units of conductance $G_0= (2
e^2/\hbar)(\ell/\pi \hbar v_F)$, where $\ell$ is the length of the
structure along the $y$ direction. Fig. \ref{fig.4} presents the
conductance versus the variation of incidence energy $E$. It is
shown that the visible kinks of the conductance due to transmission
resonances are closely related to the quasi-bound states. More
importantly, all conductance curves indicate a pronounced forbidden
region, that is, the region of almost zero conductance corresponding
to the transmission gap.

Finally, we have a brief look at the reflection. Fig. \ref{fig.5}
shows the reflection probability $R=1-T$ as the function of Fermi
wavelength $\lambda$. It is shown that the electron can perfectly
reflected by the graphene barrier. This pass-band in reflection is
analogous to Bragg reflection in optics, which is also found in
magnetic barrier in graphene \cite{Ghosh}. Actually, the Bragg-like
reflection can also be applied to select electron wavelength or
energy by the reflection window.

In summary, we investigate the transmission and reflection in 2D
monolayer graphene barrier at the non-zero incidence angle. It is
shown that the transmission gap as function of the incident energy,
which results from the evanescent waves in two cases of Klein
tunneling and classical motion, can be controlled by the incidence
angle, the height and width of potential barrier. With the
realization of the tunable potential barrier in graphene
\cite{Huard}, we hope these phenomena may lead to the potential
applications in various graphene-based electronic devices.

This work was supported in part by the National Natural Science
Foundation of China (Grant No. 60806041), the Shanghai Rising-Star
Program (Grant No. 08QA14030), the Science and Technology Commission
of Shanghai Municipal (Grant No. 08JC14097), the Shanghai
Educational Development Foundation (Grant No. 2007CG52), and the
Shanghai Leading Academic Discipline Program (Contract No. S30105).
X.C. is also supported by Juan de la Cierva Programme of Spanish
MICINN.

\end{document}